\documentclass[a4paper,12pt]{article}

\usepackage{amsmath}
\usepackage{epsfig}
\usepackage{float}
\usepackage[ngerman,english]{babel}
\usepackage[utf8]{inputenc}
\usepackage{hyperref}

\newcommand{\beq}{\begin{equation}}
\newcommand{\eeq}{\end{equation}}

\begin{document}

\begin{center}

{\Large 
	{\bf $\omega$ mesons in medium - revisited} \\
}
{\large
	\vspace{0.5cm}
	Fabian Eichstädt \footnote{e-mail: fabian.eichstaedt@theo.physik.uni-giessen.de}, Stefan Leupold und Ulrich Mosel\\
}

	\vspace{0.5cm}	
	{\it Institut für Theoretische Physik \\
	     Universität Giessen \\
	     Heinrich-Buff-Ring 16, D-35392 Giessen, Germany}

	\vspace{1cm}

{\bf Abstract}
\vspace{0.5cm}

\begin{minipage}{0.85\textwidth}
Recent experimental observation by the CBELSA/TAPS collaboration indicates a downward mass shift of the $\omega$ meson when put into a nuclear medium. Therefore it is worthwhile to take a look at theoretical approaches that can describe this mass shift. One such hadronic model was introduced by Klingl et al. \cite{Kl99}. We will demonstrate that the heavy baryon limit approximation used there has no justification and returns wrong results. The results obtained with a correct treatment of the relevant $2\pi$-channel lead to unreasonable results for the $\omega$ selfenergy in medium.
\end{minipage}

\end{center}

\section{Introduction}
In the past a lively discussion has been going on about a possible mass shift of the $\omega$-meson in a nuclear medium. While there seems to be a general agreement that the $\omega$ aquires a certain width of the order of 40-100 MeV in the medium, the matter of mass shift is not so commonly agreed on. While some groups have predicted a dropping mass, e.g. \cite{Kl99}, \cite{Kl97}, \cite{Re02}, there have also been suggestions for a rising mass \cite{DM01}, \cite{PM01}, \cite{SL06}, \cite{Zs02} or even a structure with several peaks \cite{Lu02}, \cite{Mu06}. In this context a recent experiment by the CBELSA/TAPS collaboration is of particular interest, since it is the first indication of a downward shift of the mass of the $\omega$-meson in a nuclear medium \cite{Tr05}. Since Klingl et al. \cite{Kl97} were one of the first to predict such a downward shift it is worthwhile to look into their approach in some more detail. 

In the course of the present paper the effective Lagrangian approach of \cite{Kl97} is taken and examined further. The authors combined vector meson dominance with a pseudovector pion-nucleon interaction and SU(3) flavor symmetry for coupling nucleons, pseudovector mesons and vector mesons, but no direct $\omega$-nucleon-resonance interaction was included.

In this framework the authors calculated the $\omega N$ forward-scattering amplitude and connected it to the $\omega$-selfenergy via the low density theorem. The inelastic processes $\omega N \rightarrow \pi N$ and $\omega N \rightarrow 2\pi N$, that are relevant for these calculations, were only treated at tree level.

The authors used a heavy baryon limit approximation (HBL), on behalf of which several possible diagrams at tree level were not included. In the present paper these calculations are reexamined by a relativistic calculation (still at tree level).

In the next section the underlying model for the calculations and the HBL are introduced. Section [\ref{results}] then will comprise the results of our calculations and a comparison with the previous results by Klingl et al. Finally in section [\ref{conclusion}] we draw our conclusions. For further details of the present calculations we refer to \cite{Ei06}.

\section{The model} \label{model}
In the present paper we are interested in the calculation of the $\omega$ spectral function in a nuclear medium:
\beq A_{med}(q) = -\frac{1}{\pi} \text{Im}\frac{1}{q^2 - (m_\omega^0)^2- \Pi_{vac}(q) - \Pi_{med}(q)}, \label{spectral-function} \eeq
with the bare mass $m_\omega^0$ of the $\omega$. The vacuum part of the $\omega$ selfenergy $\Pi_{vac}$ is determined by the decay $\omega \rightarrow \pi^+ \pi^0 \pi^-$ \cite{Kl96}. For the calculation of the in-medium part one can employ the low density theorem \cite{Kl97}, \cite{Lu02}, \cite{Mu06} which states that at sufficiently small density of the nuclear medium one can expand the selfenergy in orders of the density $\rho$:
\beq \Pi_{med}(\nu, \vec q = 0; \rho) = - \rho T(\nu) \label{low-density-theorem}\ , \eeq
where $T(\nu)$ is the $\omega$-nucleon forward-scattering amplitude. Note that all the calculations are made for isospin-symmetric nuclear matter at temperature $T=0$. Also, as in \cite{Kl97}, the scattered $\omega$ is taken to have $\vec q = 0$ relative to the nuclear medium. This approximation is justified by the fact that one expects to see an in-medium effect for $\omega$'s which decay inside the medium.

The model which is employed to calculate the forward-scattering amplitude is introduced in detail in \cite{Kl97}. For completeness we give the conventions and coupling constants that we used:
\begin{align*}
g_A &= 1.26, \\
g_{\rho N} &= 6.05, \\
g_{\omega N} &= 3 g_{\rho N} \\	
f_\pi &= 0.0924\ \text{GeV}, \\
\kappa_\rho &= 6.0, \\
\kappa_\omega &= 0, \\
g_{\rho\omega\pi} &= 1.2 \\
\gamma^5 &= \begin{pmatrix} -1 & 0 \\ 0 & 1 \end{pmatrix} \\	
\text{Tr} \left( \gamma^\alpha \gamma^\beta \gamma^\mu \gamma^\nu \gamma^5 \right) &= -4i\epsilon^{\alpha\beta\mu\nu}\
\end{align*}
As motivated above we are interested in the $\omega$-nucleon forward-scattering amplitude at the one-loop-level. To obtain the imaginary part of this amplitude via Cutkosky's Cutting Rules one needs as input the inelastic reactions $\omega N \rightarrow \pi N$ ($1\pi$-channel) and $\omega N \rightarrow 2\pi N$ ($2\pi$-channel). However one has to decide which of these processes should be included in the calculations. Klingl et. al \cite{Kl99}, \cite{Kl97} employed HBL to select the relevant processes for the $2\pi$-channel. Note here that in our and in Klingl's work the $1\pi$-channel remains unaffected by this limit, i.e. in contrast to the $2\pi$-channel no processes are neglected. 

\begin{figure}[H]
\hfill
\begin{minipage}[b]{.45\textwidth}
\begin{center}  
\includegraphics[width=6.5cm]{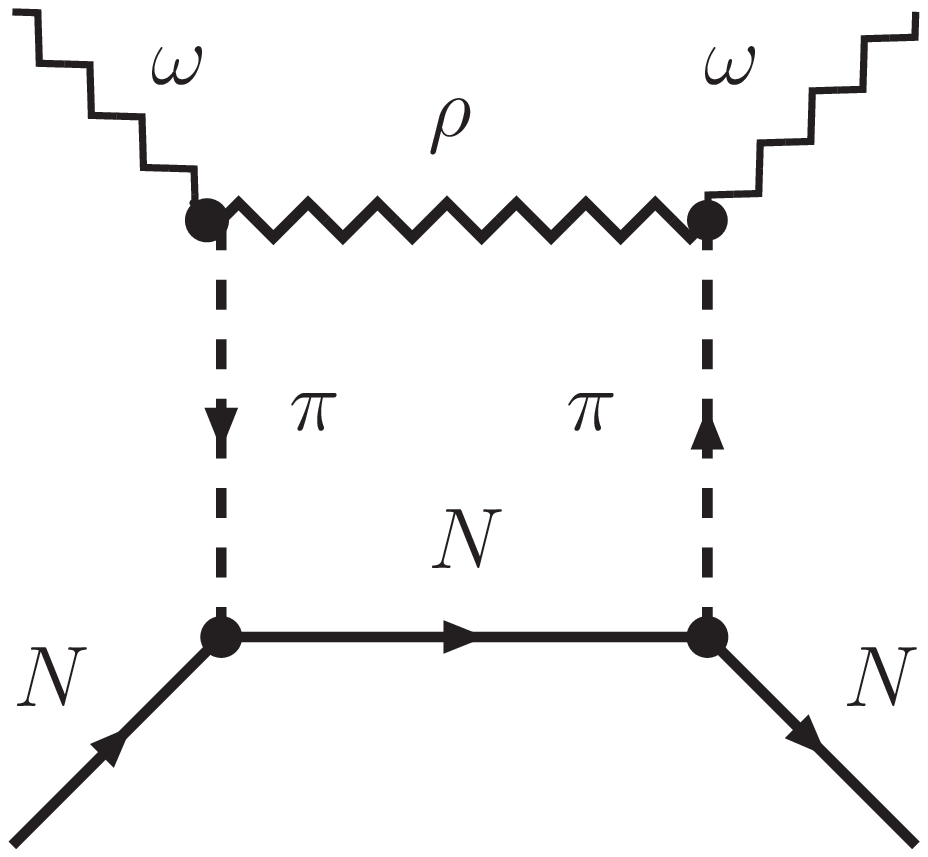}
\caption{Box diagram with intermediate $\rho$}
\label{diag-j}
\end{center}
\end{minipage}
\hfill
\begin{minipage}[b]{.45\textwidth}
\begin{center}  
\includegraphics[width=6.5cm]{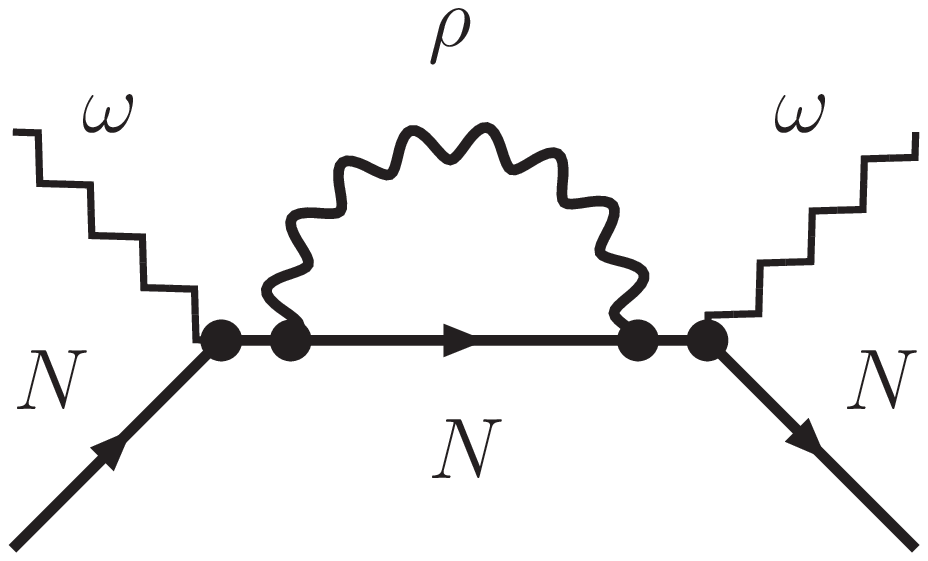}
\caption{Diagram with direct $\omega N$ coupling}
\label{diag-k}
\end{center}
\end{minipage}
\hfill
\end{figure}

Examining the HBL one finds that all processes with direct $\omega N$ coupling are suppressed. For example in this limit the diagram in figure \ref{diag-k} contributes nothing to the total scattering amplitude while the diagram in figure \ref{diag-j} becomes the most important one. Both these diagrams belong to the $2\pi$-channel since the intermediate $\rho$ meson decays into two pions.

In the next section we will compare the results of the calculations with and without HBL.

\section{Results} \label{results}
Since the $\omega N \rightarrow \pi N$ decay channel is not influenced by HBL and since our calculations agree with the results by Klingl et al. we do not look into this channel further at this point. It should be stressed that the amplitude $\omega N \rightarrow \pi N$ is more or less fixed by the measurable and measured back reaction \cite{Fr97}. This is in contrast to the reaction $\omega N \leftrightarrow \rho N$ which is barely constrained by data. It is the latter reaction, however which decides about the in-medium mass shift of the $\omega$ \cite{Kl99}.

\begin{figure}[H]
\centering
\includegraphics[width=12cm]{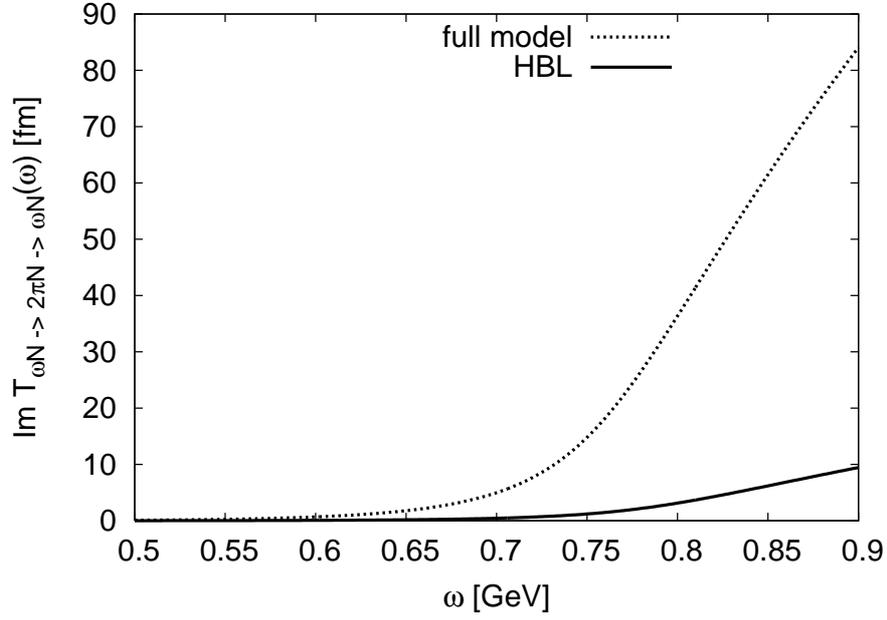}
\caption{Comparison of the imaginary part of the amplitudes for the $2\pi$ decay channel, calculations with the full model and with HBL.}
\label{2pi-channel-imag}
\end{figure}

In contrast to the one pion channel we find considerable differences in the $2\pi$ decay channel. This is shown in figure \ref{2pi-channel-imag} where calculations of the forward-scattering amplitude $\text{Im}\ T$ with the full model and with HBL are compared. It is obvious that the results are one order of magnitude larger in the former case than in the latter one. This is in striking contrast to the results of \cite{Kl97} where the diagram in figure \ref{diag-j} is identified as the most important one. We thus have to conclude already at this point that HBL is unjustified for the processes considered here and leads to grossly incorrect results.

\begin{figure}[H]
\centering
\includegraphics[width=12cm]{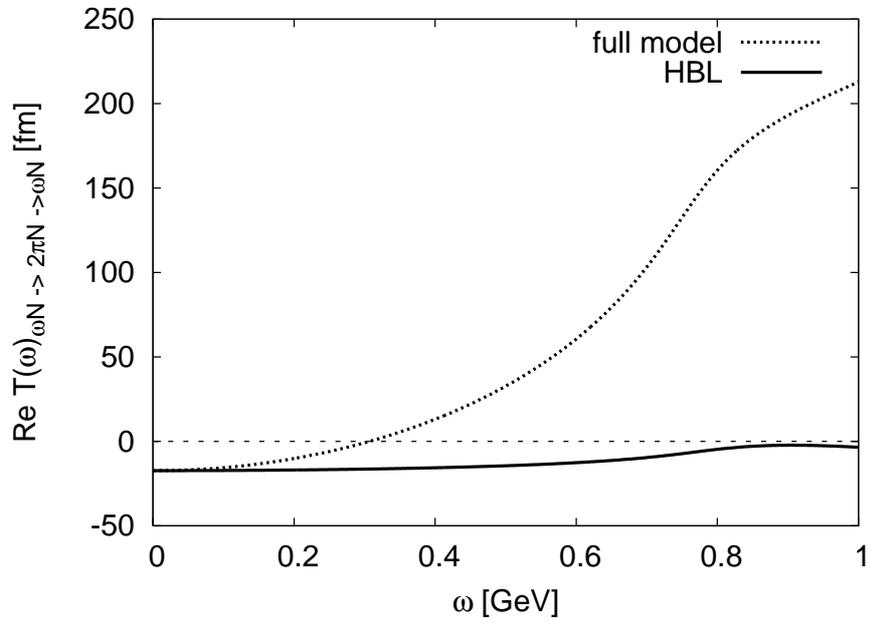}
\caption{Comparison of the real part of the amplitudes for the $2\pi$ decay channel, calculations with the full model and with HBL.}
\label{2pi-channel-real}
\end{figure}

Since the real part of the amplitudes are obtained from the imaginary part through a dispersion relation this drastic behavior also translates into the real part. As one can see in figure \ref{2pi-channel-real} the result for the real part of the $\omega N \rightarrow 2\pi N$ scattering amplitude in the full model even has opposite sign to the result with HBL, which means attraction in the former and repulsion in the latter case. Note here that the overall attraction that is found by Klingl et al. \cite{Kl99} is based on the fact that in their calculations even more processes (e.g. $\omega N \rightarrow 3\pi N$) are included. Note here also that the real part starts off at a negative value because a once subtracted dispersion relation was used (as in \cite{Kl97}) and the subtraction constant is negative.

\begin{figure}[H]
\hspace{1.85cm}
\includegraphics[width=9cm]{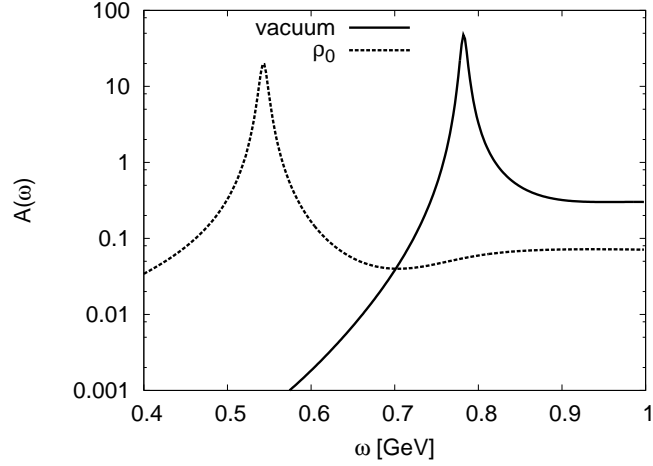}
\caption{Spectral function of the $\omega$ meson in the vacuum and at normal nuclear density.}
\label{omega-spec-func}
\end{figure}

Finally we show our resulting in-medium spectral function of the $\omega$ (where HBL was not employed) in figure \ref{omega-spec-func}.  Note that in the medium the peak is shifted to 544 MeV which is due to the large effects of a relativistic, full treatment of the imaginary and real parts of the amplitudes obtained in the present model.

\begin{figure}[H]
\centering
\includegraphics[width=9cm]{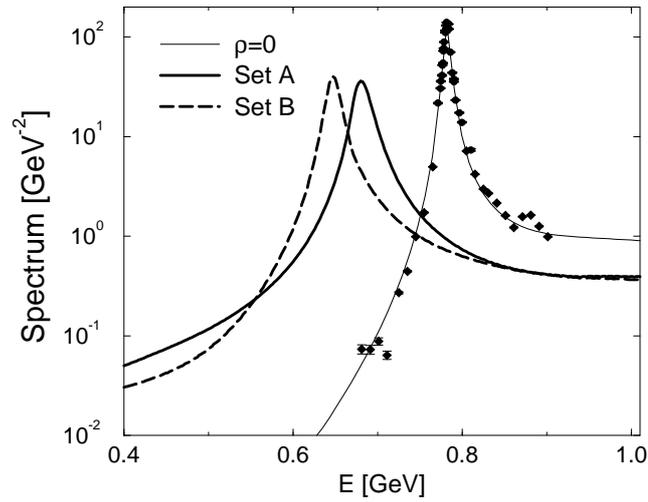}
\caption{Spectral function (compare Set A to our results) of the $\omega$ meson as obtained by Klingl et al. \cite{Kl99} by employing HBL. (from \cite{Kl99})}
\label{klingl-spec-func}
\end{figure}

This has to be compared with the results obtained by Klingl et al. \cite{Kl99} which are shown in figure \ref{klingl-spec-func}. Since Klingl et al. find an in-medium peak at about 620 MeV it is obvious that in the relativistic calculation the physical picture changes drastically.

\section{Conclusion} \label{conclusion}
This work was triggered by recent experimental indications \cite{Tr05} of a significant lowering of the mass of the $\omega$ meson in nuclear matter. A prediction describing this situation had been made by Klingl et al. \cite{Kl99}, \cite{Kl97} and thus their approach was well worth to be studied further. They connected the $\omega N$ forward-scattering amplitude with the in-medium $\omega$ selfenergy using the low density theorem and thus were able to make a prediction for the spectral function of the $\omega$ meson in nuclear matter. In this calculation HBL was extensively used. Since other calculations yield different results for the $\omega$ mass shift in medium we have reexamined the calculations of Klingl et al. using the identical Lagrangian.

Figures \ref{2pi-channel-imag} and \ref{2pi-channel-real} demonstrate, that diagrams with a direct $\omega N$ interaction give the greatest contribution to the imaginary part of the scattering amplitude, compared to which e.g. the diagram in figure \ref{diag-j} can basically be neglected. In contrast in \cite{Kl97} this diagram is found to be one of the largest contributions whereas other diagrams are not considered further, because of HBL. In the present work these diagrams were calculated fully relativistically. The comparison of our results and the results by Klingl et al. in the last section shows that by including these diagrams the results change dramatically. This makes the use of HBL unjustified.

In the last section our results for the in-medium $\omega$ mass were presented, indicating a decrease of around 240 MeV from the vacuum mass. This decrease appears to be very high, especially in light of recent experiments \cite{Tr05}, which suggest an in-medium mass of 722 MeV at $0.6 \rho_0$ nuclear density. This drastic behavior can be traced back to the relativistic treatment of the $2\pi$-channel in the model of \cite{Kl97}, since the $1\pi$-channel can basically be fit to data (as shown in \cite{Kl99}). The large contributions of this $2\pi$-channel in a correct (relativistic) treatment of the assumed Lagrangian indicate a problem of this low energy hadronic model to realistically describe the $\omega$ in the medium.

This problem might partially be based on the fact that all the inelastic processes $\omega N \rightarrow \pi N$ and $\omega N \rightarrow 2\pi N$ are only treated at tree level. Here an improved calculation is needed, which incorporates coupled-channels and rescattering, e.g. a Bethe-Salpeter \cite{Lu02} or a K-matrix approach \cite{Fe98}, \cite{Pe02}, \cite{Sh04}.

\newpage
The authors acknowledge discussions with Norbert Kaiser and Wolfram Weise. This work has been supported by DFG.

\end{document}